\documentclass[12pt]{article}
\usepackage{amsmath}
\usepackage{graphicx}
\usepackage{epsfig} 
\usepackage{amssymb,amsbsy}
\usepackage[dvipsnames]{color}

\def\be{\begin{equation}}
\def\ee{\end{equation}}
\def\ba{\begin{eqnarray}}
\def\ea{\end{eqnarray}}

\begin{document}

\title{ Notes on counter-orbiting globular clusters in the Milky Way }
\author{V.~Yankelevich\footnote{Email: vik-yank@yandex.ru} \\
 Southern Federal University, Rostov-on-Don, Russia }

\date{}

\maketitle


\begin{abstract}
It is argued that Galactic globular clusters (GC) rotating retrograde may originate from prograde GCs that change their angular momentum 
due to gravitational perturbations from the Magellanic Cloud galaxies. It is shown that those galactic GCs with orbits near the Lagrange 
point of the system ``Milky Way -- Maggellanic Clouds'' can change the sign of their angular momentum in few Gyr time scale. 
\end{abstract}

\newpage


\section{Introduction}

\noindent
The Galactic globular cluster system counts 150 clusters: 33 of them are recognized to orbit prograde, i.e. in the same direction as the halo 
stellar population of the Milky Way does, 15 of them are retrograde, i.e. counter-rotate the stellar halo, while for 
the rest the rotation direction is not identified yet \cite{ha,mars}. The origin of retrograde GCs is widely debated and remains totally 
controversial. A common understanding is that they are born outside the Milky Way and only relatively recently -- a few Gyr ago -- have been captured 
by the MW gravitation field on to confined orbits \cite{van,bo,kr}. However a possibility that they could have formed in the MW and later on have 
changed their rotation under the gravitational influence from the Andromeda galaxy cannot be totally excluded (see discussion in \cite{kr}). 
In this model the captured clusters are to differ from the galactic ones in two respects: first, they seem to have higher angular 
momenta and lie predominantly on the peripheric regions of the Milky Way, and second, in general they are expected to form a separate 
population of clusters. 

In Fig. 1 we show the distance-metallicity relation for those globular clusters with an identified orientation of their 
rotation from the Harris catalogue \cite{ha}. It is clearly seen that the distributions for prograde and retrograde are 
mostly similar and hardly can be distinguished: the retrograde clusters are spread lightly narrower and lie to a certain extent 
 {\it within} the distribution 
of the prograde ones. A very vague anti-correlation between the metallicity and the distance for the two cluster populations does not look 
different. In addition, the most distant clusters are strikingly the progrades: {\it i)} the most distant prograde cluster orbits at around 
100 kpc which is 3 times of the orbit of the most distant retrograde; and {\it ii)} the mean galactocentric radius for the prograde clusters 
$\langle R\rangle=11.76\,{\text{\rm kpc}}$, while for the retrograde clusters $\langle R\rangle=10.19\,{\text {\rm kpc}}$, which is in conflict 
to an intuitive expectation. Moreover, the most distant retrograde cluster has the galactocentric radius $R\simeq 30$ kpc, while the most 
distant prograde one has $R\simeq 100$ kpc -- half an order higher. Crosses in Fig. 1 show globular clusters with uncertain direction of 
orbital motion, and the same behavior as demonstrated by the prograde and retrograde clusters are readily observed here. Overall, from this 
point of view the retrograde clusters do not seem to differ crucially from the other clusters. The only distinguishing feature in Fig. 1 is that the 
retrograde clusters are less spread in metallicities, which might be though attributed to individual evolutionary pecularities.  

Neither differences between the progrades and retrogrades are seen in the distributions of their total (the sum of azimuthal and radial) 
velocities $\upsilon_t=\sqrt{\upsilon_{\theta}^2+\upsilon_r^2}$ as shown in 
Fig 2: both the populations show nearly equally wide distributions ranging from $|\upsilon_{\theta}|\simeq 100$ km s$^{-1}$ to 
$|\upsilon_{\theta}|\simeq 400$ km s$^{-1}$, apart from one retrograde with $\upsilon_t=550$ km s$^{-1}$ which is mostly due to 
the radial motion, nor in the other distributions, as for instance, the ``mass-radius'' and the ``mass-metallicity'' relations for 
both cluster subpopulations (Fig. 3a and 3b). Therefore, besides their reversed orbital motions the retrograde clusters are fairly 
similar to the whole population of galactic globular clusters and can hardly be 
distinguished from the other clusters.      


\begin{figure}
\begin{center}
\includegraphics*[width=0.6\textwidth]{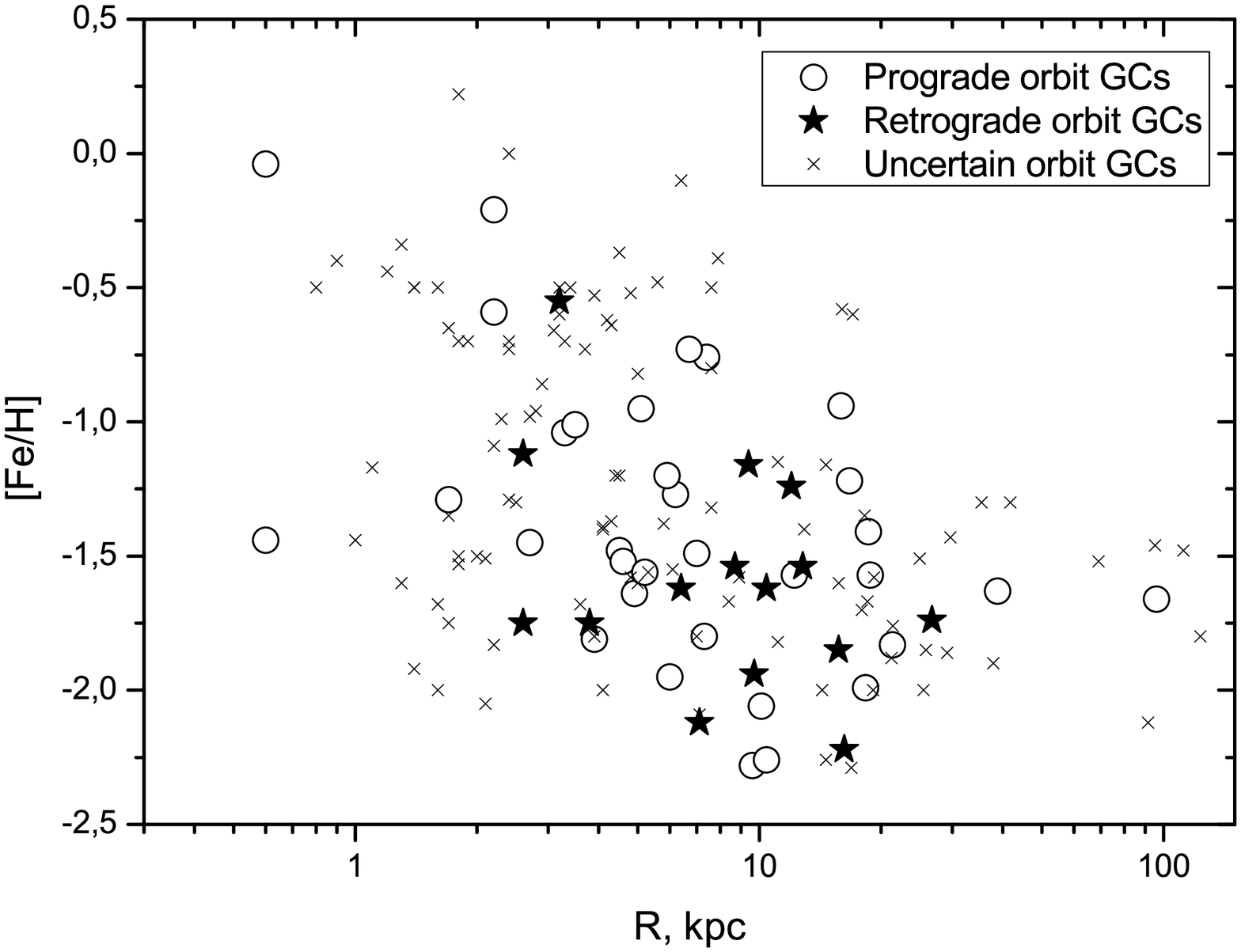}
\end{center}
\caption{
The ``distance-metallicity'' relation for prograde GCs (open circles) and retrograde GCs (asterisks); crosses show 
clusters with unidentified rotation direction.
}
\label{fig1}
\end{figure}

\begin{figure}
\begin{center}
\includegraphics[width=0.6\linewidth]{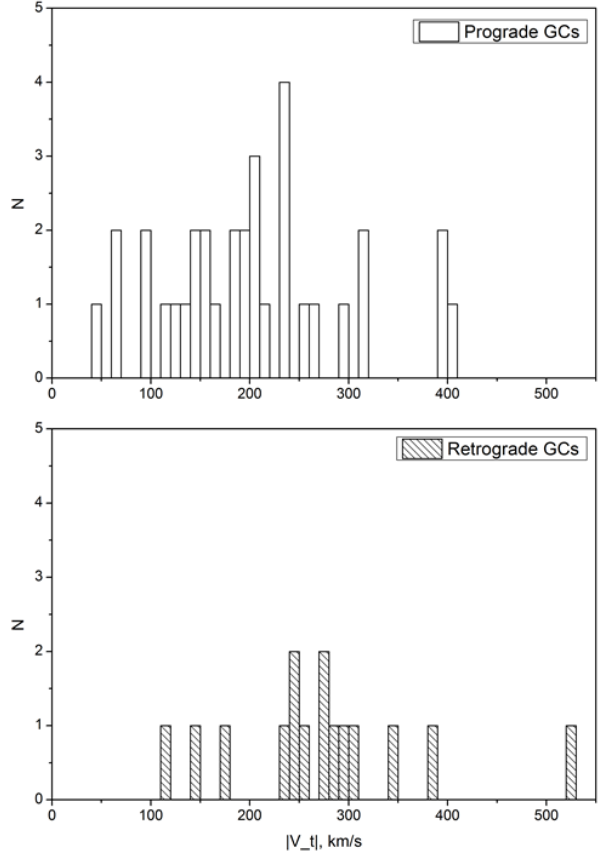}
\end{center}
\caption{ The distribution function of the modulus of total velocity $\upsilon_t=\sqrt{\upsilon_r^2+\upsilon_\theta^2}$ km s$^{-1}$ for 
for the prograde (upper) and retrograde (lower) clusters; high velocities of the retrogrades are due to their high radial velocities.   
} 
\end{figure}

\begin{figure}
\begin{center}
\includegraphics[width=0.4\linewidth]{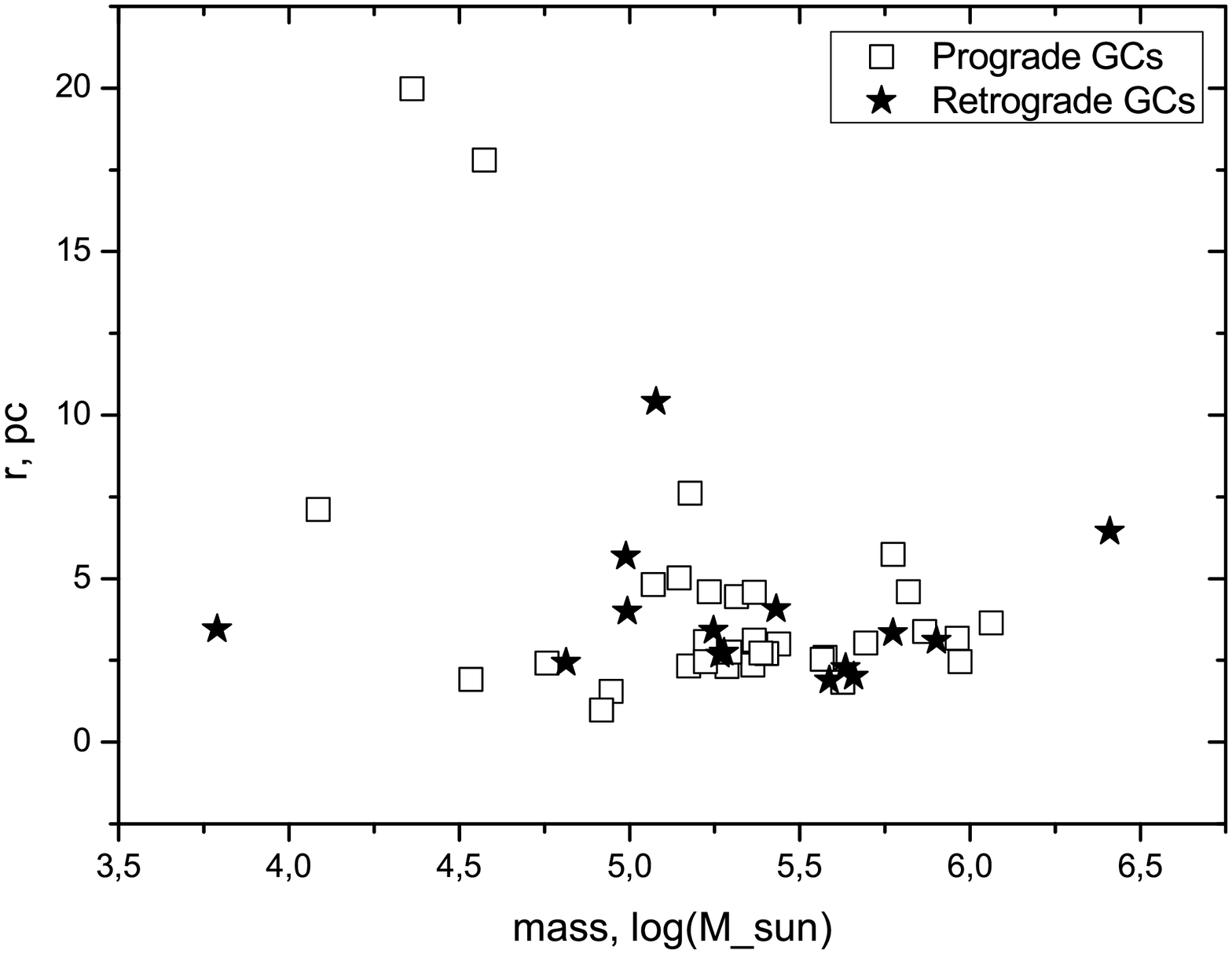}
\includegraphics[width=0.4\linewidth]{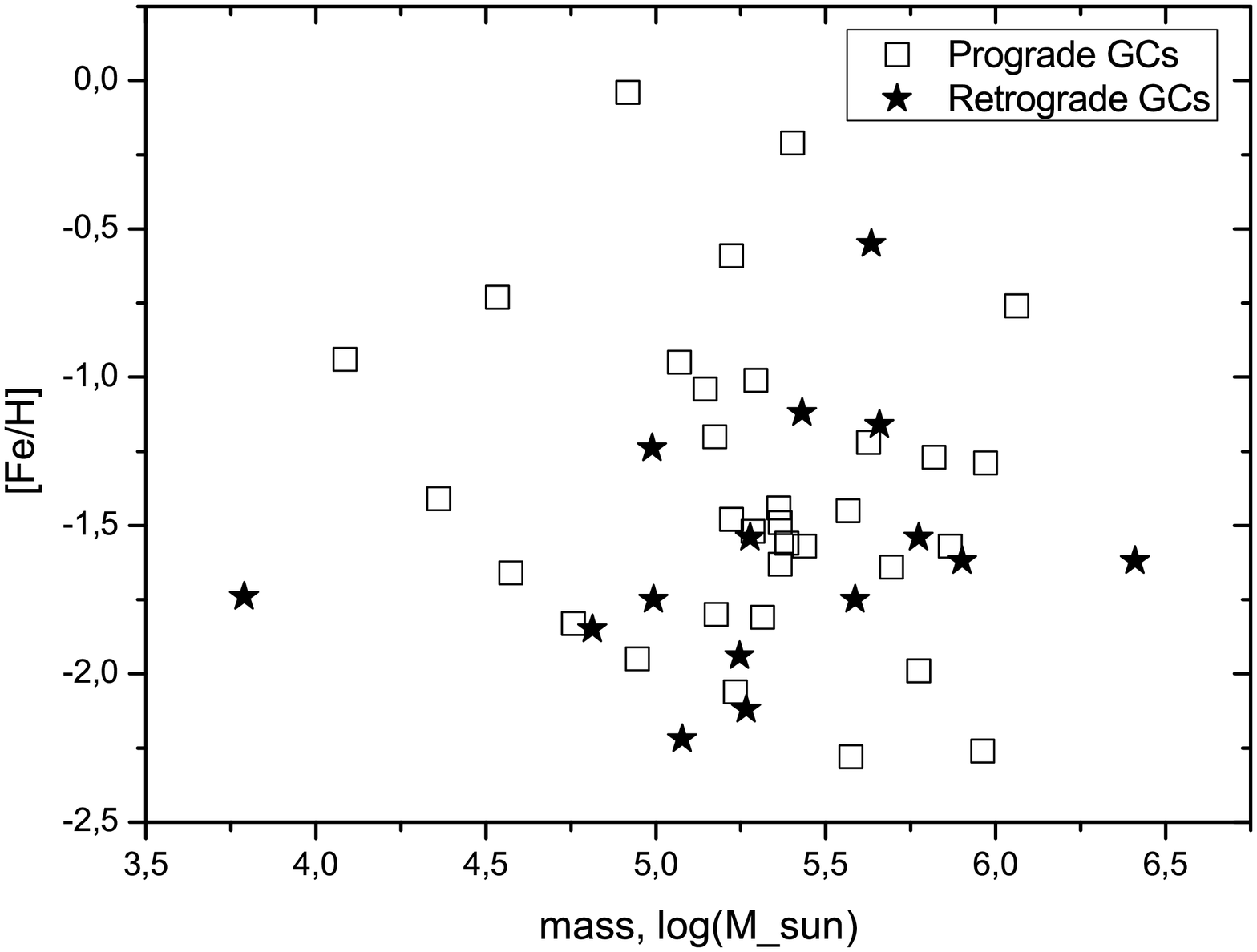}\\
\end{center}
\caption{ The ``mass-radius'' (left and ``mass-metallicity'' (right) relations for the progrades (open squares) and retrogrades (asterisk). 
} 
\end{figure}

In this brief note we demonstrate a principal possibility for galactic globular clusters to change the sign of their 
angular momentum under the action of the gravitational perturbations from the Magellanic Clouds. 


\section{Reversal of angular momentum of a globular cluster}

\noindent
Where the origin of retrograde globular clusters is concerned it is implicitly assumed that gravitational potential of the Milky Way 
is spherically (or axially) symmetric, which implies conservation of the angular momentum of globular clusters. However, as soon as the 
orbits of globular clusters may extend up to distances comparable to the separation between the Milky Way and Magellanic Clouds, the 
non-axisymmetricity of the potential and possible violation of the angular momentum conservation have to be noted and accounted for.   

In order to demonstrate the possibility for globular clusters to reverse their angular momentum, in this paper we apply a simplistic dynamical 
model of a cluster moving in the joint gravitational field of the Milky Way and the Magellanic Clouds. For the Milky Way we assume the total 
mass $M_{\rm halo}=7\times 10^{11}~M_\odot$, the halo raius $R_{\rm halo}=400$ kpc 
\cite{bo,deas,so}, and the Navarro-Frenk-White (NFW) density profile \cite{nav} 

\be
\label{eq1}
\rho(r)=\frac{\rho_0}{\frac{r}{h}(1+\frac{r}{h})^2}
\ee
where $h=12.5 \text{\rm kpc}$, $\rho_0=1.06\cdot 10^{-2} M_{\odot} {\rm pc}^{-3}$. 
The initial conditions for the positions and the velocities of globular clusters are assumed to correspond to Harris catalog. It is worth 
stressing that current estimates of the galactic mass are rather uncertain and may vary within factor 4--5 \cite{so,mc,ir}. Depending on its exact value 
dynamics of the Magellanic Clouds may change dramatically from orbiting confined gravitationally to the Milky Way to move gravitationally 
unbound \cite{besla}. Our choice here corresponds to the gravitationally bound system ``Milky Way -- Magellanic Clouds'' with the 
velocities $\upsilon_x=0$, $\upsilon_y=-219$ km s$^{-1}$, $\upsilon_z=186$ km s$^{-1}$ for the LMC, and 
$\upsilon_x=0$, $\upsilon_y=-174$ km s$^{-1}$, $\upsilon_z=173$ km s$^{-1}$ for the SMC \cite{ned}, the accepted total masses are \cite{so}: 
$M_{_{\rm LMC}}=1.6\cdot 10^{10}~M_{\odot}$, and $M_{_{\rm SMC}}=6.5\cdot 10^{9}~M_{\odot}$.


We start calculations of dynamics of a GC from a typical configuration shown in Fig. 4. 
In general, only those clusters experiencing sufficiently 
strong influence of the non-axisymmetricity of the gravitational potential 
whose initial positions and velocities do lie in a restricted box. In 
other words, only the clusters occupying a restricted volume of the initial phase space are attracted in further evolution 
to the Magellanic Clouds. Therefore, 
a number of runs covering a set of initial conditions has to be checked in order to determine the trajectories which could, in principle, 
change considerably the angular momentum in the vicinity of the Magellanic Clouds. 

We performed simulations by making use of the ordinary differential equation Runge-Kutta solver of 7-8 order.  
Each run covered 10 Gyr evolution with the accuracy $10^{-8}$. 
    

In order to demonstrate the very possibility of a reversal of the angular momentum for globular clusters, first of all, we show here a typical example for a cluster
starting from a box of the initial phase space with trajectories that attract to the Magellanic Clouds: $y=36$ kpc, $z=36$ kpc, $\upsilon_0=212$ 
km s$^{-1}$; the trajectory is depicted in Fig. 4: in this particular example the cluster reverts its momentum in 7.45 Gyr. The 
corresponding evolution of the angular momentum is shown in Fig. 5 -- the reversal occurs on a short time scale of $\sim 1$ Myr. 

\begin{figure}
\begin{center}
\includegraphics*[width=0.6\textwidth]{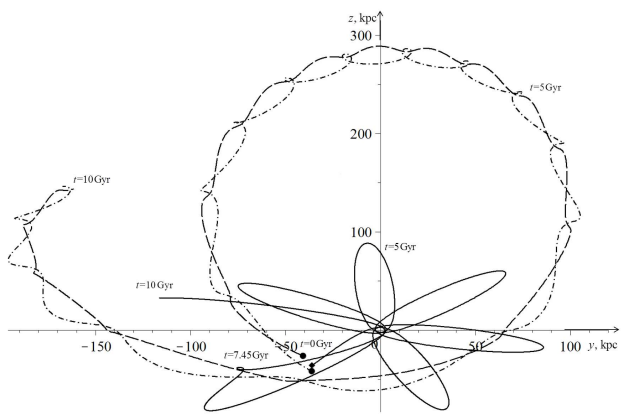}
\end{center}
\caption{
The trajectory of a GC moving in the joint field of the NFW density profile of the Milky Way and point-like potentials of LMC and SMC. 
Filled square and circles at around $y=-40$ kpc and $z=-40$ kpc correspond to the initial positions of the 
clobular cluster, and LMC and SMC galaxies, respectively; thin solid line is the trajectory of the GC, long dash line 
corresponds to the trajectory of LMC, while dash dotted line to the SMC. At $t=7.45$ Gyr the critical interaction of the GC 
with the orbiting each other SMC and LMC followed by a loop in the GC's trajectory is clearly seen. 
}
\label{fig4}
\end{figure}

\begin{figure}
\begin{center}
\includegraphics*[width=0.6\textwidth]{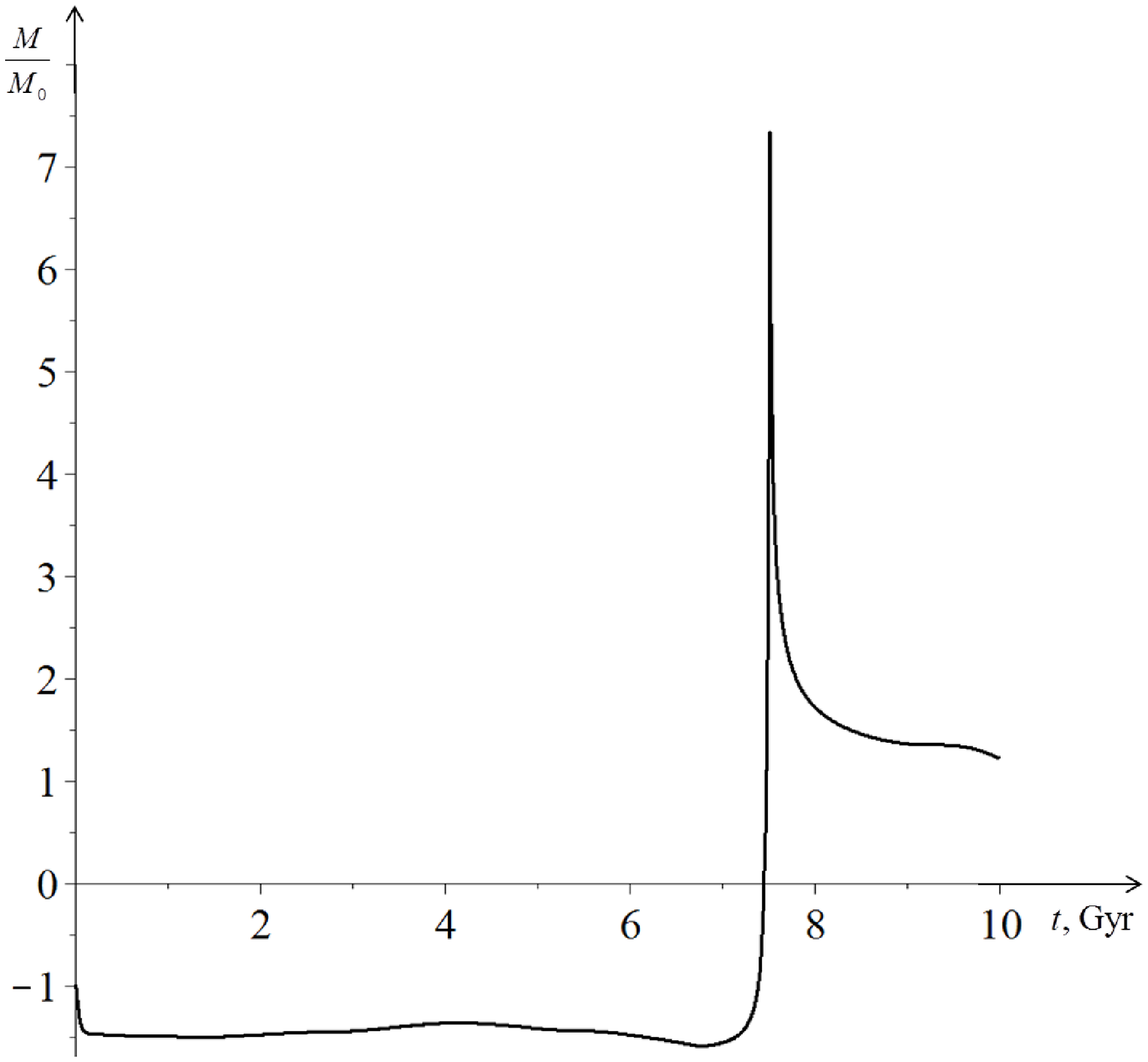}
\end{center}
\caption{
The graph of total angular momentum of GC in the Navarro-Frenk-White density profile.   
}
\label{fig6}
\end{figure}

Further, we demonstrate the quantity of GCs which potentially change angular momentum.
  A set of 3D 4 body simulations was performed. All the initial parameters of MW, LMC and SMC were the same as in the previous example. We took $10000$ test particles for GCs, where the initial distance from the galactic center is distributed according with the law $r\cdot e^{\frac{r}{6}}$. This law approximately coincides with the real distribution of known galactic clusters \cite{ha},\cite{mars}. The initial distribution of polar and azimuthal angles of GC is even. All the initial velocities except $\upsilon_{tet}$ are distributed by the normal law with dispersion $130$ km/s, for $\upsilon_{tet}$ we used the absolute value of the normal distribution with dispersion $130$ km/s. Thus all the GCs are prograde in the beginning. Each run covered $10$ Gyr evolution with the accuracy $10^{-8}$.
In the results there is about $3-4\%$ of the GCs which become retrograde. Thus we demonstrate that the gravitational interactions in our system can change the direction of rotation of a GC though the number of such clusters is relatively small.


\section{Conclusion}

\noindent
We have presented arguments favoring a new concept of the origin of the retrograde globular clusters. We have found that when a cluster starts 
from a selected volume of the initial phase space it falls into the area of attraction to the Magellanic Clouds, where strong non-axisymmetric 
violations of gravitational potential may change the cluster orbital angular momentum within the Hubble time. 
As a result, some of the clusters from this selected 
volume of the initial phase space may reverse the momentum, and turn from normal prograde rotation to the retrograde. We therefore argue that 
at least some of the retrograde clusters might have originated from the galactic population of globular clusters having been formed in 
the initial phase space with trajectories attractive to the Magellanic Clouds. Note, however, that this our scenario works only if the 
Magellanic Cloud galaxies are gravitationally confined to the Milky Way galaxy. 

\section{Acknowledgments}

\noindent
I thank prof. Yu. A. Shchekinov for formulation of the problem and guidance, prof. 
V. A. Marsakov for discussions and providing his processed Harris catalog, Douglas Heggie for useful discussions. 
I acknowledge support from the ``Dynasty'' Foundation,  the Ministry of Education and Sciences of RF (project coded 14.A18.21.0787). 
Part of this work has been done while I was visiting the Henri Poincare Institute.


\end{document}